\documentstyle[11pt,newpasp,twoside,epsf,fleqn]{article}

\def\edcomment#1{\iffalse\marginpar{\raggedright\sl#1\/}\else\relax\fi}
\reversemarginpar

\begin{document}
\vspace*{1cm}
\title{Chances for earth-like planets and life\\ around metal-poor stars}
 \author{Hans Zinnecker}
\affil{Astrophysikalisches Institut Potsdam, An der Sternwarte 16,\\
14482 Potsdam, Germany}
%\author{Co-Author}
%\affil{The Name of My Institution, The Full Address of My Institution}

\begin{abstract}
We discuss the difficulties of forming earth-like planets
in metal-poor environments, such as those prevailing in the
Galactic halo (Pop II), the Magellanic Clouds, and the
early universe. We suggest that, with less heavy elements
available, terrestrial planets will be smaller size and
lower mass than in our solar system (solar metallicity).
Such planets may not be able to sustain life as we know it.
Therefore, the chances of very old lifeforms in the
universe are slim, and a threshold metallicty (1/2 solar?)
may exist
for life to originate on large enough earth-like planets.
\end{abstract}

\section{Intro}

We do not know whether life is widespread in the Galaxy and the universe.
Many factors have to conspire to form habitable earth-like planets, with
the right conditions to sustain  a biosphere (see the book
``Rare Earth'' by Ward and Brownlee). One of these factors seems to be
a certain amount of heavy elements and metals (C, N, O; Mg, Si, Fe;
radioactive elements and also phosphor central to the RNA/DNA world).
Therefore the question
arises if there is a critical metallicity that has to reached before
life can originate. For example, life-bearing planets probably need
enough liquid water (containing oxygen) at their surface and a
sufficiently strong magnetic field generated in their iron core
to shield any incipient lifeforms from lethal external UV radiation
and energetic cosmic rays. Both the amount of oxygen (to form water)
and the amount of iron (to form a sizable core) are likely to be
significantly reduced in a metal-poor galactic environment, such as
in the Galactic halo or in globular clusters. We refer to these
conditions as Population II conditions, but broaden them to include
star and planet formation environments in the Magellanic Clouds,
with heavy element abundances between 1/4 solar (LMC) to 1/10 solar
(SMC).

In this short note, we argue that planets like Earth
are unlikely to form under conditions as metal-poor as the SMC or
even the LMC (regions with dust to gas mass ratios as low as
1\,:\,1000 or 1\,:\,2000), because with such a reduced dust content any
terrestrial planets that may form in circumstellar disks around
solar-mass stars (Beckwith \& Sargent 1996)
will be significantly smaller in size
than Earth (perhaps Mercury or Mars size), and hence unable to retain
enough of an atmosphere, among other things. Thus we suspect that there
is nobody up there in the Galactic halo
or in the Magellanic Clouds
looking down on us in the Galactic disk!

\section{Metallicity Dependence for Earth-like Planet Formation}

The Earth has  been formed by the collisions of some 10 Mars-sized
protoplanets, which themselves have been formed through runaway and
oligarchic growth of so-called planetesimals (solid bodies
with sizes of order of 5\,km and masses of the order of 10$^{18}$\,g);
see Kokubo \& Ida (2000) for runaway and oligarchic growth and
Hayashi et al. (1985) for the initial masses of planetesimals at
an orbital radius of 1\,AU. If we now make the assumption that
in a metal-poor environment the dust mass density scales linearly with
the metallicity (a reasonable assumption it seems, 
see Bouchet et al. 1985), we can extrapolate
the formalism of Kokubo \& Ida (2000) to the case of a metal-poor
circumstellar disk 
with a reduced dust surface mass density, proportional to 
metallicity ($Z/Z_\odot$).
The result is that the final masses of protoplanets
for 1/4 and 1/10 solar metallicity will be scaled down to
0.02 earth masses and 0.005 earth masses, respectively
(instead of 0.16 earth masses for solar metallicity), as
the mass of a protoplanet ($M_{proto}$) scales with the
3/2 power of the dust surface mass density ($\Sigma$):

\begin{equation}
M_{proto}\sim \Sigma ^{3/2}\sim \left(Z/Z_\odot\right)^{3/2}
\end{equation}

The mass of a protoplanet after oligarchic growth and in
a circular orbit is essentially the integrated
surface mass density of dust in a ring whose width ($w$) is given
by the Hill (Roche) criterion:

\begin{equation}
w \approx 10\,\left( \frac{M_{proto}}{M_*}\right)^{1/3}a
\end{equation}

\noindent
where $M_*$ is the mass of the central star (1\,M$_\odot$) and a is the 
orbital radius (1\,AU). We find $w\approx0.1AU$ for $Z/Z_\odot=0.1$.
It follows that in the 1/4 or 1/10 solar metallicity cases
the orbital spacing of protoplanets is twice or three times
as tight as in the solar metallicity case.
The growth timescale for protoplanets at $a=1AU$ will be
about 100,000 to 250,000\,yr in the two metal-poor cases,
inversely proportional to the surface mass density, i.e.
the metallicity.
The protoplanet system formed by oligarchic growth will become
orbitally unstable on longer timescales (eccentricity pumping
and orbit crossing) and will form more massive bodies by
collisional accretion. How massive will the most massive terrestrial planet
be (the equivalent of the Earth)? Long-term N-body simulations
(such as those by Wetherill \& Stewart 1993)
would need to be carried out, but we can estimate the outcome.
We need 5\,--\,20 sticky collisions (ignoring fragmentation) to grow
the protoplanets to the size and mass of Mars (10\,\% earth mass).
The collision time between the protoplanets turns out to be
of the order of a few times 10$^8$\,years, assuming a velocity
dispersion of the order of the orbital velocity (30\,km/s) and
a geometric cross-section for collisions (protoplanet sizes
of order 1000\,km).
This implies these
protoplanets may just make it into Mars-like objects in
half a Hubble time, but they won't make it into Earth-sized
planets; there is not enough time and there is not enough
material within the effective annular region for bigger planets
to form.

\section{A threshold metallicity for life to begin?}

In the previous section, we have found that terrestrial planet
formation is a sensitive function of metallicity. It seems we
need a metallicity rather close to solar (at least half solar)
for planets the size and mass of Earth to form. This is due to
the non-linear dependence of mass growth and timescales on the
surface mass density of dust in the habitable zone (near 1\,AU).
Decreasing the surface mass density of solid material has
a dramatic effect on the final outcome: rather than Earth-like
planets we get asteroids and gravel, not suitable for life.

Life depends in many ways on the size and mass of a planet.
Firstly, as mentioned before, mass $M$ and radius $R$ determine
the gravity
of planet \linebreak ($g=GM/R^2$) which grows roughly proportional to size.
Therefore bigger planets can better retain more of an atmosphere,
which is crucial for our life (oxygen, ozone, carbon dioxide,
water and rain), and may be crucial for other life-forms, too.
Secondly, if too small, a planet will not have enough of an
iron core to generate
a substantial magnetic field by the dynamo effect, which is
required to shield the surface from cosmic ray bombardment.
Thirdly, if
the planet is too small, there is not enough heat generated
by the radioactive elements and by the collisional build-up
that the planet can sustain volcanic activity and plate
tectonics for a long time, both instrumental for the
carbondioxide cycle
and a stable greenhouse effect. In short, if the planet is
too small, it cools off quickly and becomes a dead world.

So far we did not discuss the issue of the formation of
a Jupiter-like planet under metal-poor conditions. We know
that the presence of Jupiter in the solar system is a big
advantage for protecting the earth from late impacts.
If Jupiter could not form under metal-poor conditions,
because the seed rocky core does not
grow large enough to attract and to accrete the gas (Mizuno 1980),
we may not have the benign environment and shelter
that enables life to blossom -- in case low-metallicity
conditions somehow manage to produce an Earth after all.

Finally, we don't know at this point, what would be the
the mix of chemical elements on a metal-poor (Pop II) planet.
Perhaps gas phase studies of the Magellanic Clouds could
help. In any case, oxygen, nitrogen, carbon, and phosphor
are needed to make amino-acids, the basic ingredients
for life to start. Liquid water with all its wonderful
properties is also needed to get going. Will there be
relatively less water on a planet that started out with
metal-poor initial conditions? Less rain? Will the
chemistry and mineralogy develop differently?
These are interesting questions
to study in the future. Today 
there are at least some indications that the 
metallicity of stars with giant planets seems
to exceed a critical threshold (Gonzales et al. 2001, 
Santos et al. 2001).

\section{Caveats and observational tests}

We note that the metallicity bias against Earth-like planet
formation can be circumvented if circumstellar disks around Pop II
stars are correspondingly smaller at lower metallicity
(smaller implies lower disk angular momentum).
In this case the smaller dust-to-gas mass ratio is compensated
by distributing the dust mass over a smaller disk, thus keeping
dust surface mass density as high as in the solar metallicity case.
[It is the dust surface mass density
that enters all the equations and causes the diversity of
planetary systems (cf. Kornet et al. 2001)].
By squeezing the disk (by a factor
of 2\,--\,3 for the LMC and SMC metallicity conditions), Earth-like
planets around metal-poor stars should be able to form.
Perhaps Jupiter-like planets, too, although by a different process:
namely by direct gravitational instability of
a gas disk, a process which is independent of metallicity and 
the dust content of a protostellar disk (Boss 2002). 

In this context, the search
for Jupiter-like planets around thousands of stars of the Pop II
Galactic globular cluster Omega Cen may be particularly worthwhile,
as this cluster unlike any other globular cluster contains stars
with a range of metallicities, thus allowing a new test of the
metallicity dependence of planet formation (K. Freeman, personal
communication). The Omega Cen cluster (1/30 solar in the mean)
is a better choice than
the much denser, constant metallicity (1/5 solar) 47 Tuc cluster,
where such a search was unsuccessful (Gilliland et al. 2000)
and where theoretical expectations were low anyway due to the
high stellar density in the cluster (Bonnell et al. 2001).

We also note that, although there are no very old, metal-poor
Pop II 1\,M$_\odot$ stars on the Main Sequence any more, many slightly
lower-mass Main Sequence stars (0.8\,M$_\odot$) of old age and
low-metallicity composition exist in globular clusters
(see Baraffe et al. 1997
for models of metal-poor, low-mass stars).
Pop II stars of mass below 0.8\,M$_\odot$ have not yet evolved
to red giants and thus would not yet have swallowed their
inner planets, should these exist
(cf. Sackmann et al. 1993, Schr\"oder et al. 2001).
Therefore, life could in principle exist around these stars
with stable conditions. Moreover, a metal-poor star of 0.8\,M$_\odot$
and $Z=0.1 Z_\odot$
is about as bright as a solar metallicity star
(cf. Baraffe et al. 1997, their Tables 2\,--\,5), and their
habitable zones could be similar, with the caveat that the metal-poor star
generates a lot more UV radiation than the solar-metallicity star.
In conclusion:
Stellar evolution would allow early Pop II lifeforms to survive,
if they ever got off to a start. Thus in principle some life
in the universe could be very old, if it has not destroyed itself.

\section{Life - early on in the cosmic history?}

How early in the cosmic history can earth-like planets form, how
early can life originate?
Lineweaver (2001) wrote an interesting article about the metallicity
selection effect retarding the onset of life in the cosmos and hence
the number of habitable Earth-like planets. However, he simply assumed
a probability for Earth-like planet formation proportional to metallicity
and does not take into account the effect of a threshold metallicity,
as proposed in this paper. The difference is that in our picture, no
suitable planets form until a certain degree of chemical enrichment
has happened, while in Lineweaver's model earth-like planets can form
in very metal-poor gas, very early on in the cosmic history, albeit
with a correspondingly small probability. We suggest his analysis
should be redone including our refined hypothesis of a threshold
metallicity. The difference is important, because if there is any
chance for life to somehow start in the very early universe (i.e.
in the first Gyr), it would have had some 12\,Gyr to evolve, which is
3 times more than life as we know it from planet Earth. Then the
question ``Where are they?'' would be an even more serious one.

\section*{Acknowledgement}

Thanks to M. Rozyczka and G. Wuchterl for helpful discussions.
Thanks to MWFK Brandenburg for a travel grant (HSP-N programme).
Thanks to Carol and Ray for bringing us all together on
Hamilton Island. Thanks to M. Zadnik for being such a nice
room-mate. Thanks to M. Burton for riding the jet-ski with me.
Thanks to Ch. Lineweaver and F. Drake for clarifying comments.
And thanks to R.\,D. Cannon for his post-Symposium hospitality
in Sydney,
including an excursion to Canberra (Mt. Stromlo Observatory).

\end{document}